\documentclass[%
 amsmath,amssymb,
 reprint,%
]{revtex4}

\usepackage{graphicx}
\usepackage{dcolumn}
\usepackage{bm}

\usepackage[utf8]{inputenc}
\usepackage[T1]{fontenc}
\usepackage{mathptmx}
\usepackage{etoolbox}
\usepackage{hyperref}
\usepackage{mhchem}

\makeatletter
\def\@email#1#2{%
 \endgroup
 \patchcmd{\titleblock@produce}
  {\frontmatter@RRAPformat}
  {\frontmatter@RRAPformat{\produce@RRAP{*#1\href{mailto:#2}{#2}}}\frontmatter@RRAPformat}
  {}{}
}%
\makeatother
\begin{document}

\preprint{AIP/123-QED}

\title[High-performance amorphous superconducting rhenium films by e-beam evaporation]{High-performance amorphous superconducting rhenium films by e-beam evaporation}
\author{E.V. Tarkaeva}%
\affiliation{ 
P.N. Lebedev Physical Institute of the Russian Academy of Sciences. Leninsky Prospekt 53, 119991, Moscow, Russia
}%
\author{V.A. Ievleva}
\affiliation{Department of Physics, National Research University Higher School of Economics. Myasnitskaya Ulitsa, 20, 101000, Moscow,  Russia}
\affiliation{ 
P.N. Lebedev Physical Institute of the Russian Academy of Sciences. Leninsky Prospekt 53, 119991, Moscow, Russia
}%

\author{A.R. Prishchepa}
\affiliation{ 
P.N. Lebedev Physical Institute of the Russian Academy of Sciences. Leninsky Prospekt 53, 119991, Moscow, Russia
}%

\author{E.S. Zhukova}%
\affiliation{ 
Moscow Institute of Physics and Technology, 141700 Dolgoprudny, Russia
}%

\author{A. Terentiev}%
\affiliation{ 
Moscow Institute of Physics and Technology, 141700 Dolgoprudny, Russia
}%

\author{A.Yu. Kuntsevich}%
\email{alexkun@lebedev.ru}
\affiliation{ 
P.N. Lebedev Physical Institute of the Russian Academy of Sciences. Leninsky Prospekt 53, 119991, Moscow, Russia
}%

\date{\today}

\begin{abstract}

We present electron beam evaporation of rhenium films on room-temperature substrates. The films are shown to be amorphous and achieve a record-high critical temperature for rhenium — exceeding 7 K at the midpoint of the transition — alongside a high critical current density of 5000 A/mm² and critical fields above 10 T. Terahertz spectroscopy reveals a BCS-like character of superconductivity with a zero-temperature energy gap of approximately 2 meV and subgap optical conductivity. Despite being friable, the films are stable and compatible with lift-off processes that opens the capabilities for superconducting device applications.
\end{abstract}

\maketitle

Superconducting device technology requires thin films with an easy fabrication process, high critical temperature $T_c$ (> 4.2 K), critical field and critical current, reasonably high normal state resistance, and stability with respect to oxidation. The list of possible candidates is rather short. Among elementary metals except radioactive technetium ($T_c=8.2$ K), there are only four with $T_c>$ 4.2 K: lead ($T_c=7.2$ K), niobium ($T_c=9.25$ K), vanadium ($T_c=5.4$ K), and tantalum ($T_c=4.5$ K). All of these materials oxidize in air and cannot be referred to as technologically simple. However, some elementary materials with low bulk $T_c$ may have enhanced critical temperature in a thin film form. The most well-known example is Al (bulk $T_c=$1.1 K and films up to 6 K\cite{lamoise1975superconducting}; this value is affected by impurities and amorphization\cite{deshpande2025influence}). The increase of $T_c$ in thin films is driven by structural modifications, substrate-driven strain, and quantum effects.

 Rhenium is a superconducting noble metal that is tolerant to oxygen at room temperature and does not form a native oxide layer until 150\textcelsius{}\cite{Shabalin2014}. The bulk $T_c$ in this material in its hexagonal close-packed phase is 1.7 K \cite{ daunt1952superconductivity, hulm1957superconducting, jonas2022quantum}. The elevated critical temperature in rhenium films has been known for decades \cite{alekseevskii1967superconducting,pappas2018enhanced, sides2020grain}.The cubic phase has an elevated $T_c$ of about 3 K and sometimes forms when thin films are evaporated atop an insufficiently hot substrate. Deformations and defects are also known to enhance superconductivity in rhenium \cite{mito2016large,idczak2021influence}. There have been reports of $T_c =$ 7-8 K in amorphous rhenium films evaporated onto liquid-helium temperature substrates\cite{collver1973superconductivity}.  Recently, the growth of amorphous films by electrochemical methods was reported\cite{pappas2018enhanced, sides2020grain, huang2018electrodeposition,ahammed2023electrodeposited}. While suitable for mass production, the electrochemical methods do not fit micro- and nanoelectronics due to contamination and the impossibility of an on-chip fabrication of arbitrary-shaped film. 
 Another way to achieve amorphous rhenium films is ion bombardment\cite{ul1983superconducting} that is also poorly suited for manufacturability. Application of strain to the amorphous rhenium with the high $T_c\approx$ 6 K leads only to degradation of superconductivity \cite{ahammed2023electrodeposited}.

For electronic applications, there is a strong demand for simple vacuum evaporation or sputtering techniques.  Previously such methods led to $T_c$ in the range of 3-4 K \cite{teknowijoyo2023superconducting, frieberthauser1970electrical} for rhenium films.

\begin{figure}
\includegraphics[width=0.5\textwidth]{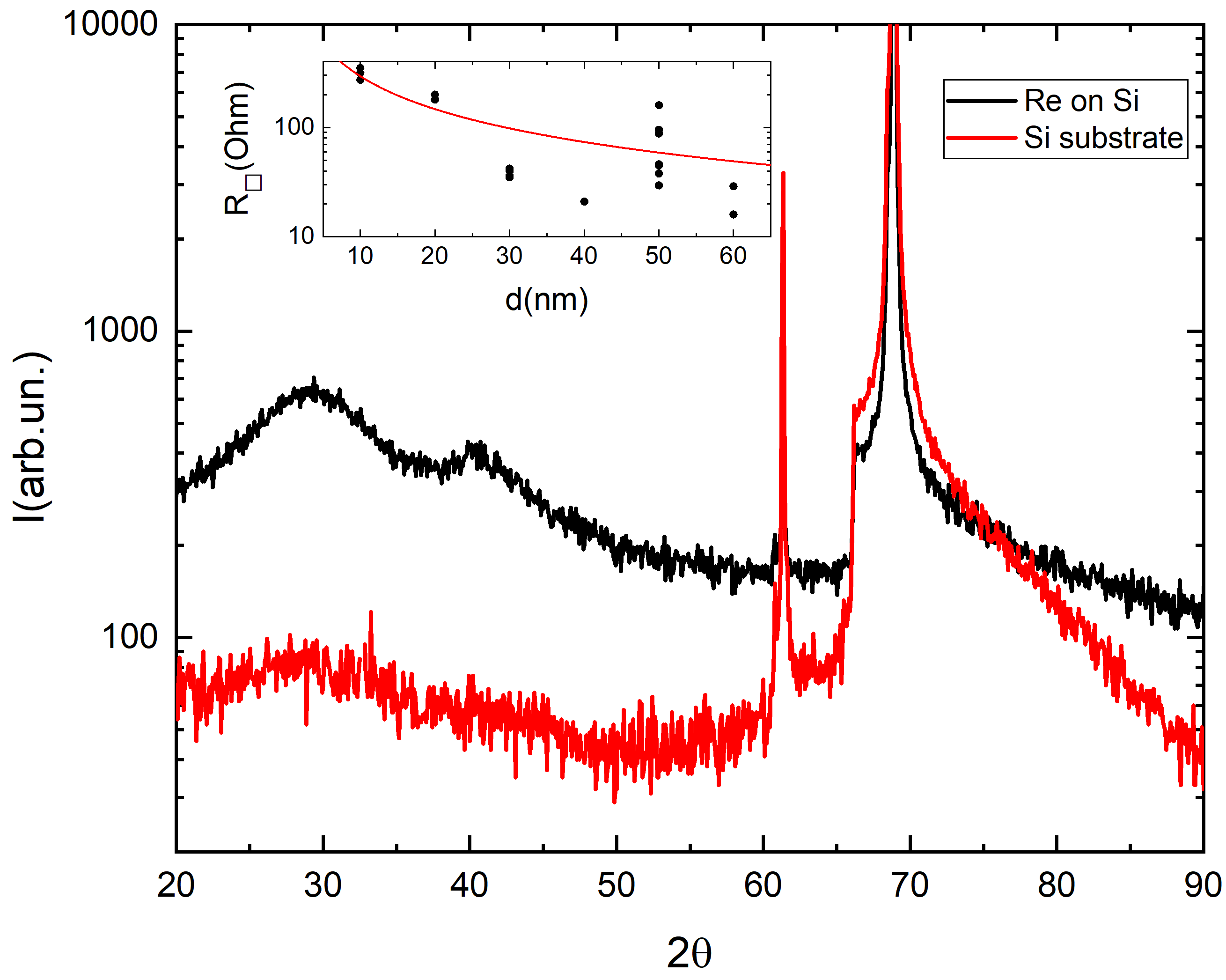}
\caption{\label{FigXRD} XRD $2\theta$ pattern of a 50-nm-thick Re film on a silicon substrate (black line) and a bare silicon substrate (red line). Inset: sheet resistance as a function of a film thickness. The red curve in the inset corresponds to an inverse proportionality fit. 
}
\end{figure}

In this paper we report electron beam evaporation growth of reproducible and technologically convenient amorphous rhenium films. For thicknesses in a range of 10-60 nm, we achieve maximum critical temperature ($T_c$>7K), critical fields of more than 10 T, and critical current density above 5000 A/mm$^2$. Terahertz measurements reveal a perfect BCS-like character of superconductivity and provide major characteristics of the superconducting state. 

\begin{figure*}
\includegraphics[width=\textwidth]{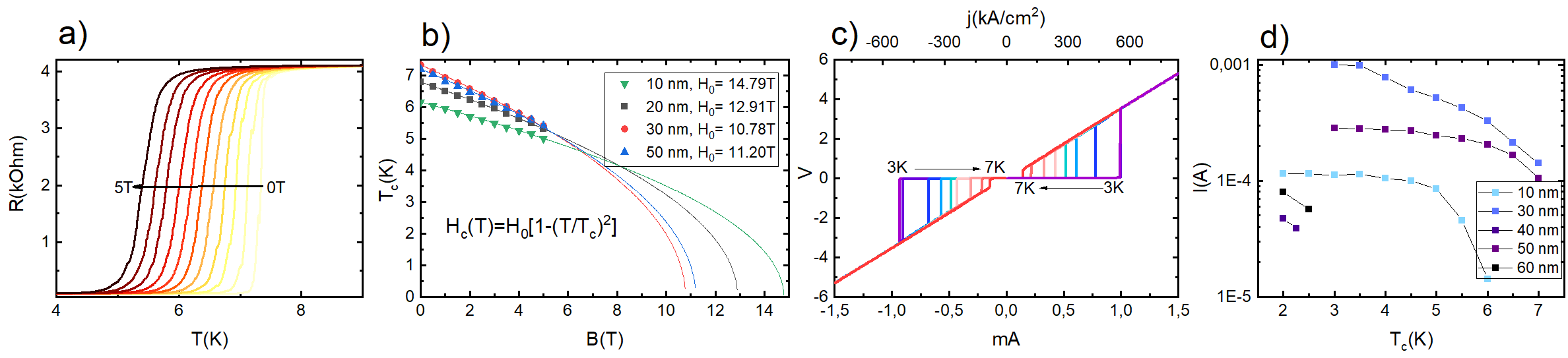} 
\caption{\label{Fig_SC} Superconducting properties of rhenium films: (a) $R(T)$ dependencies at various perpendicular magnetic fields (0, 0.5, 1, ..., 5 T) for a 30 nm Re film; (b) Temperature dependence of the critical field for several films; (c) Examples of $I-V$ characteristics for a 6 $\mu$m $\times$ 600 $\mu$m rectangular bar of a 30 nm film; (d) Temperature dependencies of the critical current density for various thicknesses of the samples.}
\end{figure*}

We grow rhenium films in the electron beam gun evaporation machine Plassys MEB 400 with a base vacuum of a $3\cdot10^{-8}$ Torr directly from a copper liner at a rate of 2 nm/min that corresponds to a current of 180 mA with a 10 kV accelerating voltage. A 99.9\% purity rhenium rod was cut into pellets by an electro-erosion and degassed by an electron beam prior to evaporation. In order to obtain amorphous films and remain compatible with the other technological processes, we intentionally avoid heating. Substrate temperature is up to 120 \textcelsius{} during an evaporation. 
For transport measurements, the films were deposited on Si substrates with a 285 nm SiO$_2$ layer, while for terahertz measurements, the films were deposited on sapphire substrates.


The amorphous nature of the films is evidenced from powder x-ray diffraction 2$\theta$ scans (see Fig.\ref{FigXRD}): no new distinct peaks appear compared to a substrate, only a shallow peaks near 2$\theta\approx 30-40^\circ$ emerge. 

Likewise, an indirect indicator of an amorphous film structure is a relatively high sheet resistance for a metallic film, that has a tendency to decrease with film thickness, as shown in the inset to Fig.\ref{FigXRD}. The ratio of the resistance at 300 K to that at 10 K is $\sim$1 in the films, which indicates strong disorder. Superconductivity is not suppressed as a film thickness decreases: 10 nm films are still superconductive. $R(T)$ dependencies for various magnetic fields perpendicular to the sample plane are shown in Fig.\ref{Fig_SC}a and are also presented in Supplementary Information, Section 1. At zero magnetic field, one can see an onset of superconductivity even above 8 K. The region where resistance drops from 90\% to 10\% of its normal state value at $B=0$ is rather narrow (a transition width is less than $100$ mK). We use 50\% criterion to identify the critical temperature. Thus determined critical fields are shown in Fig.\ref{Fig_SC}b. Extrapolation with the formula $H_c = H_{c0} \big(1 - (\frac{T_c}{T_{c0}})^2 \big)$ suggests that the value of $H_{c0}(T=0)$ considerably exceeds 10 Tesla, which gives a value of a coherence length $\xi$ of about 11-14 nm.  

Critical current was measured by a DC method in a two-terminal long rectangle geometry (6$\times$600 $\mu$m) using a Keithley 2400 source-meter and a Keithley 2000 voltmeter. Examples of IV characteristics at different temperatures are shown in Fig.\ref{Fig_SC}c (more data can be found in Supplementary Information, Section 2). In Fig.\ref{Fig_SC}d, we plot a critical current on temperature dependence for the films of different thicknesses. The critical current density value at 3 K is 5000 A/mm$^2$, which exceeds the value for Nb$_3$Sn and is just a few times less than in ultrathin NbN films.

High magnetic fields and critical currents are due to strong pinning of Abrikosov vortices. In combination with a high resistivity per square (about a hundred Ohms for a 10-20 nm thick film), it means that an operation voltage could be about several volts in rather compact devices, which is an easily measurable value. Such voltages make it possible to use Re films in single photon detectors\cite{korneeva2018optical} or bolometers\cite{shurakov2015superconducting} and also as a heated superconducting switch. Importantly, critical temperatures and currents are reproducible from cooldown to cooldown.

\begin{figure}
\includegraphics[width=0.5\textwidth]{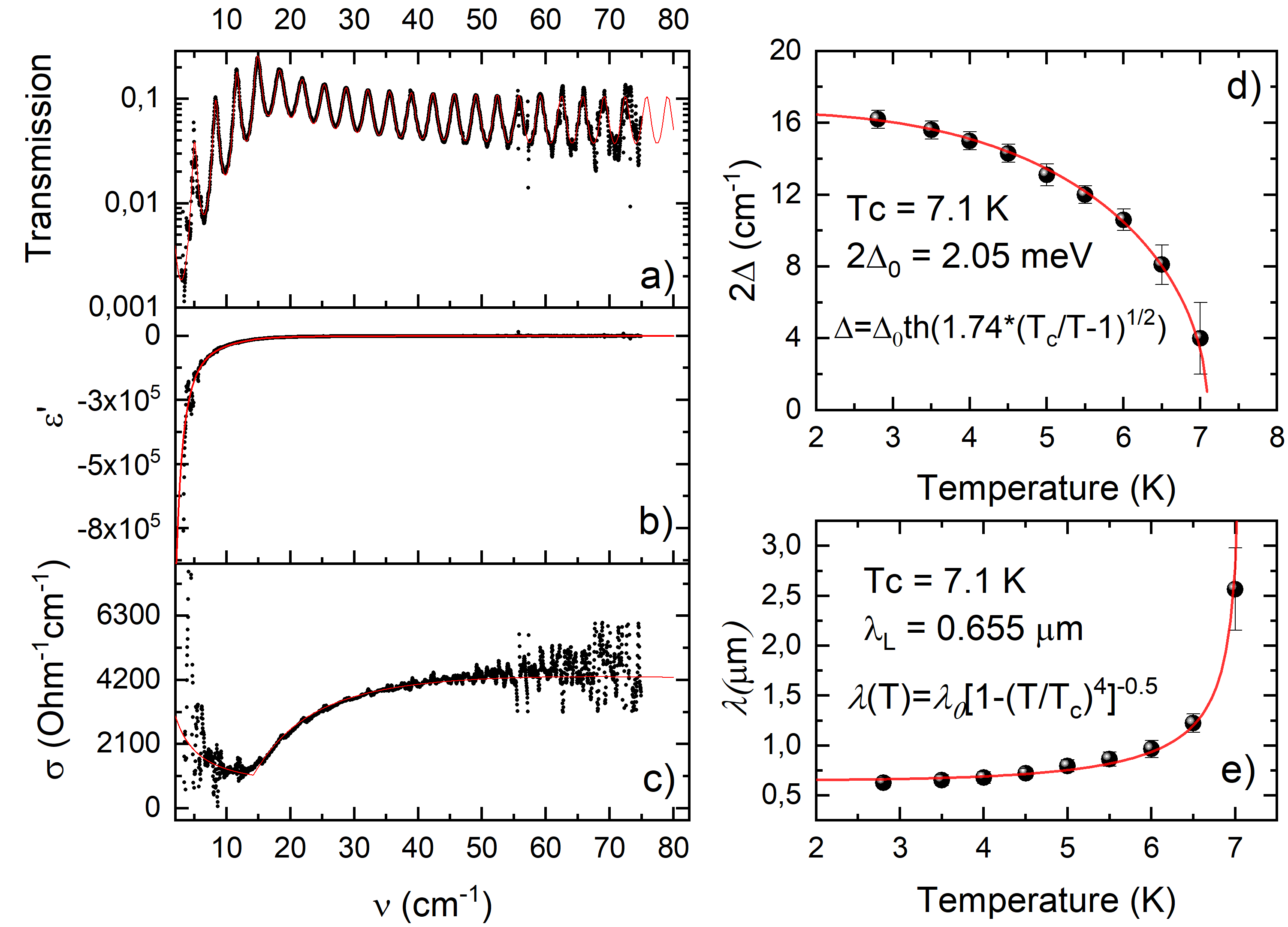}
\caption{\label{opticfigure} 
Terahertz spectroscopy results: (a) Terahertz spectrum of transmission coefficient of Re film on sapphire substrate measured at 2.5 K. (b) and (c) – terahertz spectra of real part of dielectric permittivity and AC conductivity, respectively, of Re film at T=2.5 K. Panels (d) and (e) show temperature dependencies of superconducting energy gap and London penetration depth, respectively. Red lines in panels a,b,c correspond to least-square fits with the BCS theory \cite{zimmermann1991optical} for the film properties and with the Fresnel expression for the two-layered structure (film on substrate) for transmission coefficient \cite{Born_1980, Dressel_2002}. Red lines in panels d,e - fits with empirical expression \cite{optical_CARRINGTON2003205} and with phenomenological Gorter-Cazimir two-fluid model of superconductivity \cite{Duzer_1999}, as indicated.
}
\end{figure}

 The terahertz experiments were performed at frequencies of 3--80~cm$^{-1}$ and temperatures of 2.5--300~K with a Tera K15 Menlo Systems GmbH time-domain spectrometer, which allows to measure a spectra of a complex (amplitude and phase) transmission coefficients of plane-parallel samples. Terahertz transmission (see an example in Fig. \ref{opticfigure}a) spectra were processed using a two-layer film-substrate model \cite{Born_1980, Dressel_2002}, and spectra of AC conductivity and dielectric permittivity of the film were extracted (Fig. \ref{opticfigure}c and \ref{opticfigure}b respectively). Processing the spectra with the BCS theory and with standard Fresnel expression for transmission coefficient of a two-layer system \cite{2023_Zhukova}, we extract the values of a zero-temperature superconducting gap and London penetration depth (see Fig.\ref{opticfigure}d and Fig.\ref{opticfigure}e) $2\Delta_0=2.05\ \mathrm{meV}\approx 3.5\,k_B T_c$ ($k_B$ is the Boltzmann constant) and a $\lambda_0\approx 650\ \mathrm{nm}$, which, considering $\xi=11-14$ nm, means a strong type-II character of a superconductivity. Our measurements revealed additional (relative to the SC quasiparticles contribution) below-gap AC conductivity, which is proportional to electromagnetic absorption. This can be attributed to weak links at the grain boundaries \cite{2023_Zhukova}, confirming the amorphous structure of the films.

{\bf Discussion} We have demonstrated superconducting rhenium films  with an exceptional combination of properties: chemical stability, high critical temperatures, high critical current, high critical field, and high normal state resistance. Among possible applications, we would mention detectors, kinetic inductance, and high-current elements of the superconducting integrated circuits. The chemical stability suggests that surface oxidation is impeded and allows one to expect hybrid structures without a single vacuum process, e.g., hybrid structures with 2D materials transferred atop. 

\begin{figure}[ht]
    \centering
    \includegraphics[width=0.8\linewidth]{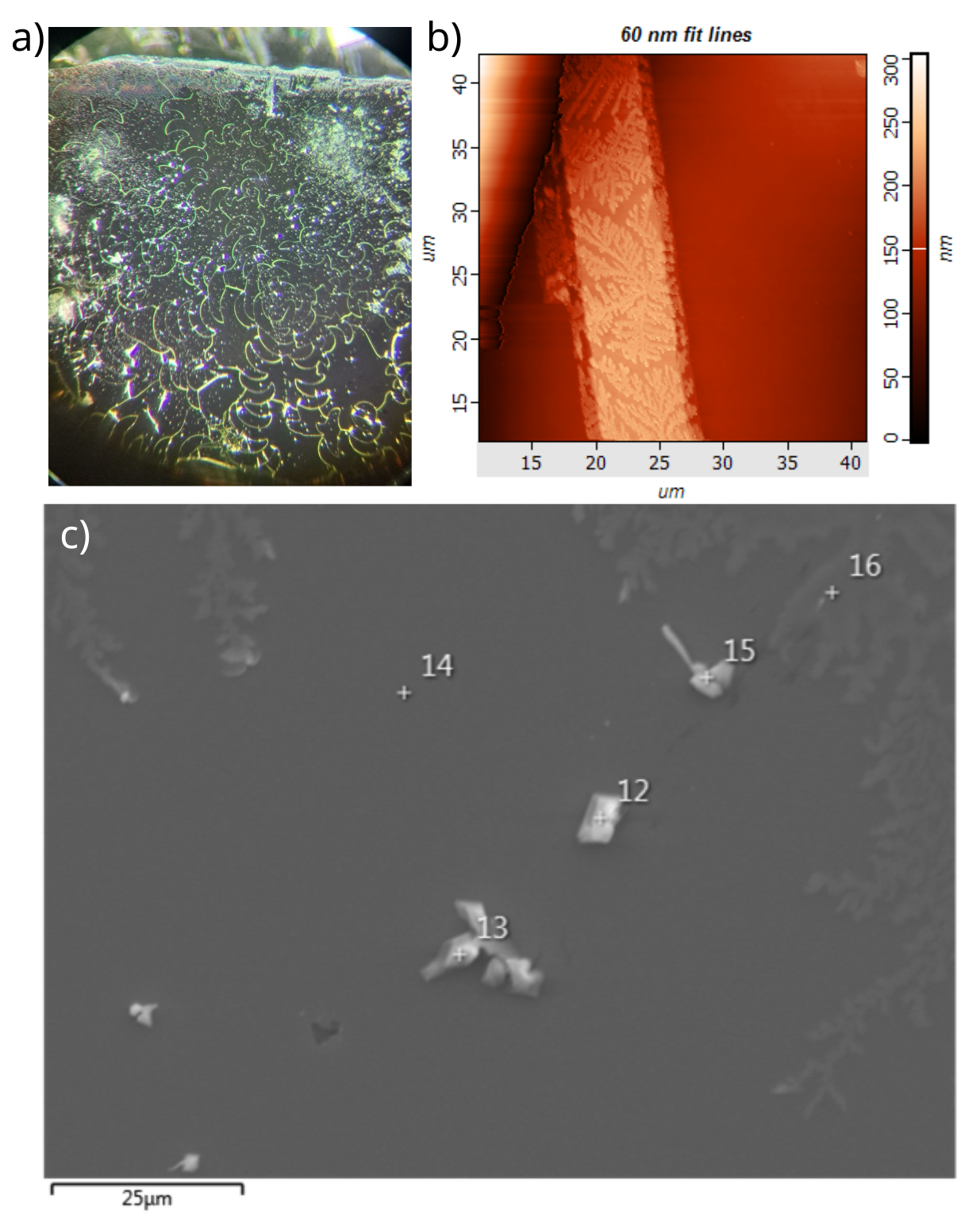}
    \caption{Surface structures formed on amorphous rhenium films: (a) Optical image of the cracks forming in cases of poor adhesion, top border of the substrate is approximately a cm long; (b) AFM scan of dendritic structures formed in several days after an evaporation; (c) SEM image of the nubbins formed one day after an evaporation, note that an energy-dispersion spectroscopy at marked points showed no signatures of elements other  than Re and a substrate material (see Supplementary Information).
    }
    \label{fig:surface_features}
\end{figure}

High melting temperature (3459 K -- second only to tungsten) makes rhenium evaporation challenging, e.g., Re is marked as poor for an e-beam evaporation on the Kurt J. Lesker website\footnote{\url{https://www.lesker.com/newweb/deposition_materials/depositionmaterials_evaporationmaterials_1.cfm?pgid=re1}}. Re pellets cannot be melted into a single piece, and evaporation occurs locally.
It is also important to avoid a contamination of Re caused by crucible material. We achieved the best results when rhenium was evaporated directly from a copper liner and systematically lower superconductive parameters when Re was evaporated from a graphite crucible.

 To investigate the role of the resist on Re films we evaporated lift-off electron beam lithography patterned rhenium bars of 0.5, 1, and 2 $\mu$m width and the same length in the same vacuum process. The superconducting critical temperatures for these bars were significantly less (about 3-4 K) than for the 6 $\mu$m width bars (see Fig.\ref{Fig_SC}). This indicates contamination sensitivity of the superconducting properties. Indeed, rhenium atoms are rather hot ($> 3000$ \textcelsius{}) and partially destroy resist walls, that leads to contamination of the rhenium edges with carbon compounds. This observation implies that the nanometer-scale lithographical route with Re should be based on inorganic shadow mask lift-off.  

Though rhenium is also one of the hardest metals in its crystalline phases, its amorphous films are not: thin films of rhenium can occasionally be scratched down to the substrate. Moreover, top layers of atoms may demonstrate migration and formation of on-surface structures (see Fig. \ref{fig:surface_features}). Previous investigations have confirmed similar surface texture effects \cite{ul1983superconducting}. 

We observed various structural formations on the film surfaces, with their specific morphology. Cracks could be formed in case of poor adhesion (see example in Fig.\ref{fig:surface_features}a). We did not obtain them systematically on well cleaned and not overheated substrates. All reported results were obtained with the films deposited without any adhesion layer. We additionally tested films with a 3-10 nm titanium underlayer, which did not affect transport properties but improved adhesion -- particularly for large-area deposition. 

The other surface structures are dendrites (Fig.\ref{fig:surface_features}b) and nubbins (Fig.\ref{fig:surface_features}c)). Their formation depends crucially on deposition parameters and lithographic processing details. The as-deposited films exhibited no immediate surface degradation and maintained structural integrity for at least 4-6 hours under ambient conditions. Following approximately 24 hours of air exposure, dendritic structures or nubbins may form, which were completely removable by rinsing with isopropanol and water. Compositional analysis (EDS, see Supplementary Information, section 3) confirmed these features to consist of rhenium, suggesting surface diffusion and reorganization of Re atoms rather than external contamination. Anyway, these structures do not affect the superconducting properties of the film bulk.

Our high-T$_c$ stable amorphous rhenium films may seem to contradict the results reported in Refs.\cite{frieberthauser1970electrical, HAQ1982119} where it was stated that films grown below $\approx$400-600\textcelsius{} were unstable and revealed cracks. At temperatures above 600\textcelsius{}, rhenium films are in a rigid crystalline form, and are, therefore, stable.  The above mentioned cracks, we believe, result from the thermal compression of a brittle amorphous phase. The films start to reveal cracks in a few hours after cooldown from growth temperature to room temperature due to a thermal expansion coefficients difference of the film and the substrate. In our case, the substrate temperature during evaporation is low enough, so that the strain is reduced and the films survive the cooling process. Deposition onto cryogenic substrates similarly to Ref. \cite{collver1973superconductivity} of course produces amorphous films. Such films, however, would be much less stable and, hence, less technology-friendly due to the inevitable adsorption of water and gases onto the substrate. Thus we are within a rather narrow temperature range in which the film adheres to the substrate without being strained by thermal expansion. We believe that unintentional substrate overheating did not allow to obtain such high-$T_c$ rhenium films previously.

\textbf{Conclusions} We demonstrate electron beam evaporation of rhenium on SiO$_{2}$ and sapphire substrates without additional heating. In the resulting thin amorphous rhenium films, a critical transition temperature exceeds 7 K, critical fields reach 10 T, and critical currents reach 5000 A/mm$^2$. 
Terahertz spectroscopy revealed a zero-temperature superconducting gap $2\Delta_0=2.05$ meV with the ratio $2\Delta_0/(k_B T_c)=3.5$, corresponding to the BCS weak coupling regime. Zero-temperature London penetration depth of 655 nm, together with the coherence length of 11-14 nm, indicate type-II superconductivity.
Transport characteristics in rhenium films are reproducible and stable over time, despite a certain migration of near-surface atoms leading to formation of textures. The demonstrated properties of the films make them promising for applications in superconducting electronic devices.




\nocite{*}
\providecommand{\noopsort}[1]{}\providecommand{\singleletter}[1]{#1}%

\end{document}